\documentclass[aps,prb,reprint,superscriptaddress,longbibliography]{revtex4-1}

\usepackage{graphicx}
\usepackage{epstopdf}
\usepackage{amsmath}
\usepackage{amsfonts}
\usepackage{amssymb}
\usepackage[a4paper=true,pagebackref=false]{hyperref}
\usepackage{color}
\usepackage{bm}

\begin{document}

\title{Probing spatial extent of topological surface states by weak antilocalization experiments }

\author{K.~Dybko}\email{dybko@ifpan.edu.pl} 
\affiliation{Institute of Physics, Polish Academy of Sciences, PL-02668 Warsaw, Poland}
\affiliation{International Research Centre MagTop, Institute of Physics, Polish Academy
	of Sciences, Aleja Lotnikow 32/46, PL-02668 Warsaw, Poland}

\author{G. P.~Mazur} 
\affiliation{International Research Centre MagTop, Institute of Physics, Polish Academy
	of Sciences, Aleja Lotnikow 32/46, PL-02668 Warsaw, Poland}

\author{W.~Wo{\l}kanowicz} \affiliation{Institute of Physics, Polish Academy of Sciences, PL-02668 Warsaw, Poland}

\author{M.~Szot} \affiliation{Institute of Physics, Polish Academy of Sciences, PL-02668 Warsaw, Poland}

\author{P.~Dziawa} \affiliation{Institute of Physics, Polish Academy of Sciences, PL-02668 Warsaw, Poland}

\author{J.Z.~Domagala} \affiliation{Institute of Physics, Polish Academy of Sciences, PL-02668 Warsaw, Poland}

\author{M.~Wiater}

\affiliation{International Research Centre MagTop, Institute of Physics, Polish Academy
	of Sciences, Aleja Lotnikow 32/46, PL-02668 Warsaw, Poland}

\author{T.~Wojtowicz}

\affiliation{International Research Centre MagTop, Institute of Physics, Polish Academy
	of Sciences, Aleja Lotnikow 32/46, PL-02668 Warsaw, Poland}

\author{G.~Grabecki} \affiliation{Institute of Physics, Polish Academy of Sciences, PL-02668 Warsaw, Poland}

\author{T.~Story} \affiliation{Institute of Physics, Polish Academy of Sciences, PL-02668 Warsaw, Poland}

\begin{abstract}
Weak antilocalization measurements has become a standard tool for studying quantum coherent transport in topological materials. It is often used to extract information about number of conducting channels and dephasing length of topological surface states.  We study thin films of prototypical topological crystalline insulator SnTe. To access microscopic characteristic of these states we employ a model developed by Tkachov and Hankiewicz, [\textcolor{blue}{Physical Review B 84, 035444}]. Using this model the spatial decay of the topological states is obtained from measurements of quantum corrections to the conductivity in perpendicular and parallel configurations of the magnetic field. Within this model we find interaction between two topological boundaries which results in scaling of the spatial decay with the film thickness. We attribute this behavior to bulk reservoir which mediates interactions by scattering events without phase breaking of topological carriers. 
\end{abstract}

\maketitle

{\it Introduction.}
The existence of topological gapless boundary states is characteristic for topological insulators (TI's)  protected by time reversal symmetry as well as for topological crystalline insulators (TCI's) governed by specific crystalline symmetry of unit cell.\cite{Molen2013i,Ando2015,Fu2010,Hsieh2012,Dziawa2012,Xu2012,Tanaka2012} 
In three dimensions the boundary states form topological surface states (TSS) with defined spin chirality. This implies their $\pi$-Berry's phase \cite{Hsieh2009} and thus robustness against elastic back-scattering. In high field magnetotransport experiments the Berry's phase of TSS manifests as an additional phase shift in Shubnikov-de Haas quantum oscillations.\cite{Dybko2017} In the low magnetic field however, the Berry's phase is responsible for destructive interference of two time reversal quantum-mechanical paths around closed loop. Due to exact quantization of Berry's phase to $\pi$ there is no other possibility than this destructive interference which leads to negative quantum correction  to resistivity at zero magnetic field.  Small magnetic field destroys phase coherence leading to positive magnetoresistance called weak antilocalization (WAL). The observation of this effect is often the first indication that we deal with topologically nontrivial material.  To date, all experiments on WAL in TI's or TCI's have been focused on measuring angular dependence of the effect to prove its 2D origin. Less attention has been addressed to the field in-plane configuration.
\begin{figure}[b]
	\includegraphics[width=0.5\textwidth]{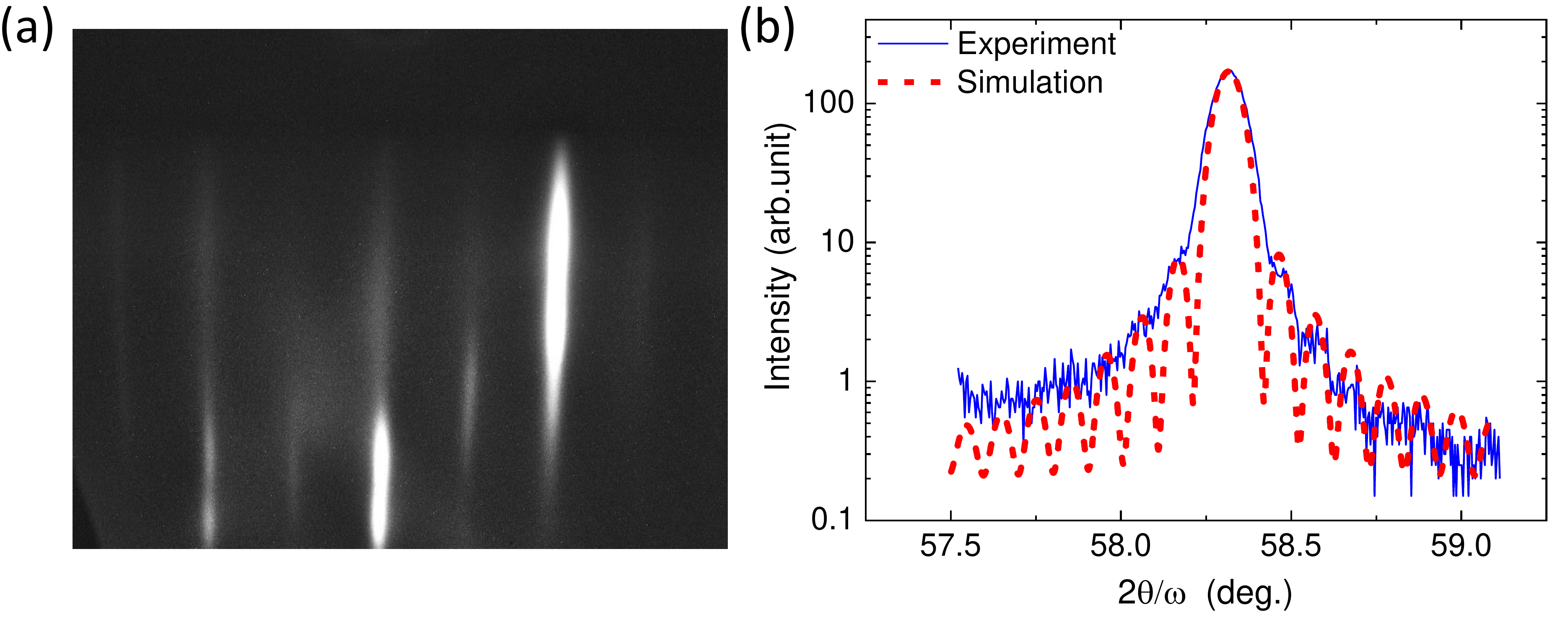}
	\caption{(a) RHEED patterns recorded during growth of the 100\,nm thick SnTe layer. (b) 004 reflection measured as a function of $2\theta/\omega$ (blue), thickness oscillations simulation (red). }
\end{figure}
A comprehensive theory for intrinsically spin--orbit coupled materials is presented in the Ref.~\onlinecite{iordanskii1994}. Commonly, the simplified Hikami--Larkin--Nagaoka (HLN) \cite{HLN1980} formalism is used for description of the experimental data in magnetic field  perpendicular to the layer. It leads to the determination of the dephasing length and  total number of conducting channels as a function of the sample thickness or gate bias voltage  in many papers devoted to Bismuth based TI's.\cite{Kim2011,Sacksteder2013}  Analogous treatment was used in IV-VI group TCI's, but the number of experiments is very limited. Already in first report  \cite{Assaf2014} the essential role of coupling between top and bottom TSS is underlined and analyzed in thick SnTe samples (thickness $\geq$ 200\,nm) grown on BaF$_2$ (001) substrates.  Similar observations reported in Refs. \onlinecite{Akiyama2016} and \onlinecite{Ishikawa2016} concerned the SnTe samples grown on   BaF$_2$ (111) and CdTe (111) substrates, respectively.
\begin{figure*}[t]
	\includegraphics[width=\textwidth]{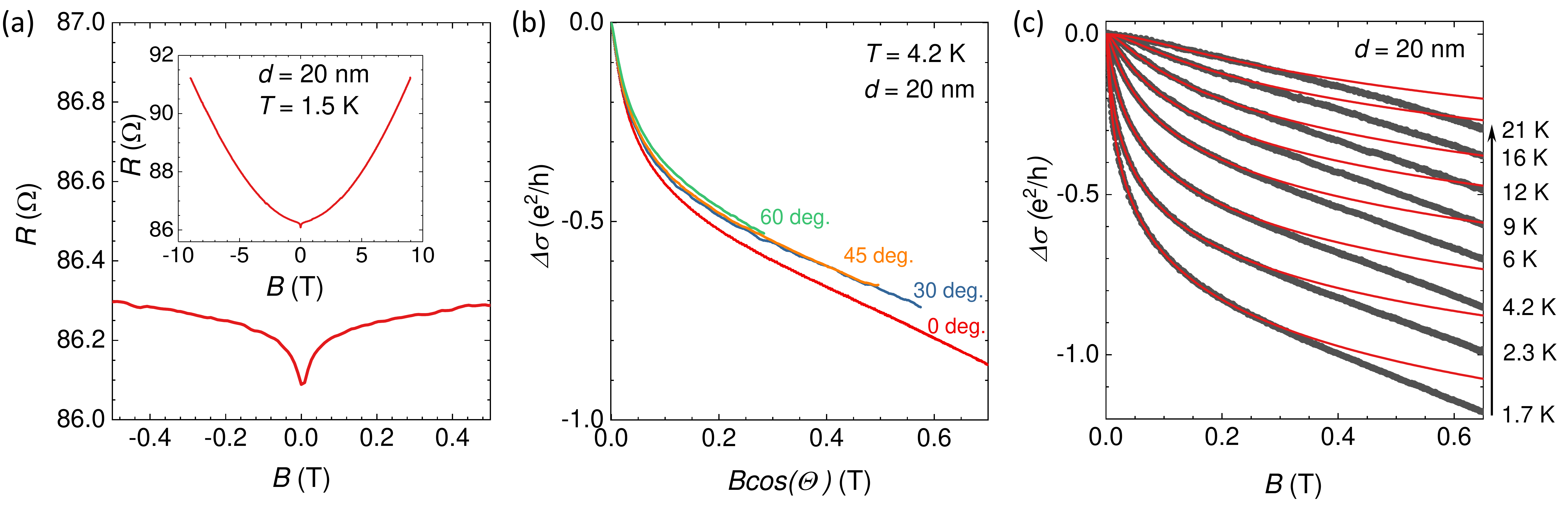}
	\caption{Representative  data set for 20\,nm thick SnTe layer (a) Raw resistance data: main panel -- low magnetic field region, inset -- whole field range, which demonstrates parabolic classical background resistance of bulk (non--topological) states (b) magnetoconductivity measured in tilted fields; $\theta$ is an angle between magnetic field direction and normal to the sample plane (c) magnetoconductivity in perpendicular magnetic field orientation measured at different temperatures; thick lines - experiment, thin lines -- theoretical fit to Eq.~(\ref{OP}).}
\end{figure*}
In this Rapid Communication we report on growth and weak antilocalization comprehensive study in SnTe as a function of layer thickness, which enabled us to determine the spatial extent of topological surface states. We employ Tkachov and Hankiewicz model \cite{Tkachov2011,Tkachov2013} of WAL developed simultaneously for two configurations of the magnetic field with respect to layer: perpendicular and parallel. The formula for perpendicular configuration is essentially identical to simplified formula of HLN.\cite{HLN1980}  Parallel field dependence involves except for the dephasing length, the additional parameter describing finite thickness of topological surface states. Such a treatment of effective thickness was already considered in early works on interference corrections to conductivity of 2DEG systems,\cite{Maekawa1981,Altshuler1981}  but now has found implementation in making complete description of weak antilocalization of TSS. Consequently, the experimental data on weak antilocalization collected in two aforementioned magnetic field configurations provide the information about the effective thickness of TSS.

{\it Experimental.} The SnTe thin layers  were grown by molecular beam epitaxy on (001) oriented CdTe(4\,$\mu$m)//GaAs substrates. All the films were grown at 350\,$^{o}$C. Finally the SnTe layers were capped with 100\,nm thick CdTe barrier. CdTe cap layer was deposited at 290\,$^{o}$C. Carrier density is estimated from Hall measurement at the room temperature and equals\\ p $\approx$ 2 $\pm$ 0.5 $\times$ 10$^{20}$\,cm$^{-3}$ for all investigated samples. The structural quality of the samples was controlled {\em in-situ} by reflection high energy electron diffraction (RHEED) and {\em ex-situ} by High Resolution X-ray Diffraction method (HRXRD)  (Figs.~1(a)  and 1(b) respectively). Both the substrate and the buffer were examined in detail to determine the possible effect on the growth of the SnTe layer. The tetragonal deformation of the lattice unit was found in the CdTe buffer layer. Biaxial compression along interface GaAs-CdTe  causes deformations of roughly $2.5 \times 10^{- 4}$ ($\varepsilon _{\rm XX}= \varepsilon _{\rm YY}= \varepsilon _{ \rm CdTe\| } = [a-a_{\rm Relax}]/ a_{\rm Relax}$ , $a$ -- in-plane lattice parameter, $a_{\rm Relax}$ - lattice parameter of not deformed CdTe unit). In the SnTe layer, tensile deformation along the CdTe interface was found. This is caused by SnTe--CdTe lattice parameter mismatch ($a_{\rm SnTe} = 6.32\,{\rm\AA }, a_{\rm CdTe} = 6.48\,\rm\AA $) or by linear thermal expansion coefficient $\alpha$  differences (for the SnTe $\alpha$ is over 3 times larger compared to CdTe).\cite{Springholz2003} The deformation of SnTe elemental cell along interface is about 0.1\%.
Quality of CdTe/SnTe interface, is revealed in thickness oscillations, (see Fig.~1(b), blue curve) recorded for reflection 004 as a function of $2\theta/\omega$. The simulation curve (marked red on Fig~1(b)) is calculated to estimate thickness oscillations characteristic of a thin (in this case 98\,nm) layer.

The samples for transport measurements of SnTe layer thickness $d$ = 10, 20, 40 and 100 nm were cleaved in the rectangular shape of width $W$ = 1.5\,mm$\div$2\,mm and length 8\,mm$\div$10\,mm. The Au/Ti metallic contacts were e-beam evaporated in a standard six probe Hall-bar type configuration.

Figure 2 (a) shows raw data of longitudinal resistance in low field range and  the inset presents the whole available field range of 9 Tesla collected at 1.5\,K. The positive low field  magnetoresistance is superimposed onto quadratic background magnetoresistance of bulk carriers. The small positive magnetoresistance we assign to topological surface states. The conductivity of 2D TSS in the whole manuscript is assumed to be $\sigma^{\rm TSS}_{\rm xx}=L/W (1/R^{\rm tot}_{\rm xx})$, where L is the distance between voltage probes. It is justified as long as low magnetic field region is considered, where  $\sigma_{\rm xy} \ll   \sigma_{\rm xx}$ holds and $\sigma^{\rm bulk}_{\rm xx} (B)$ is nearly constant.

The weak antilocalization phenomenon, in principle, could also be due to bulk states in systems without inversion center where Dresselhaus spin-orbit interaction is present. Also, in uncapped samples the inversion layer on the surface might be formed  inducing structure inversion asymmetry (triangular type quantum well) with Bychkov-Rashba spin-orbit interaction.\cite{Pfeffer1995}  Both scenarios appear unlikely in the studied system. First, SnTe is of rocksalt \textit{fcc} structure having inversion center at each atom position. Second, our SnTe films are surrounded by CdTe barriers on both sides, which prevents them from adsorption of foreign adatoms on the surface and consequently from the formation of inversion/accumulation layer.

\begin{figure}[tb]
	\includegraphics[width=\columnwidth,trim={0cm 1.5cm 0cm 0cm}]{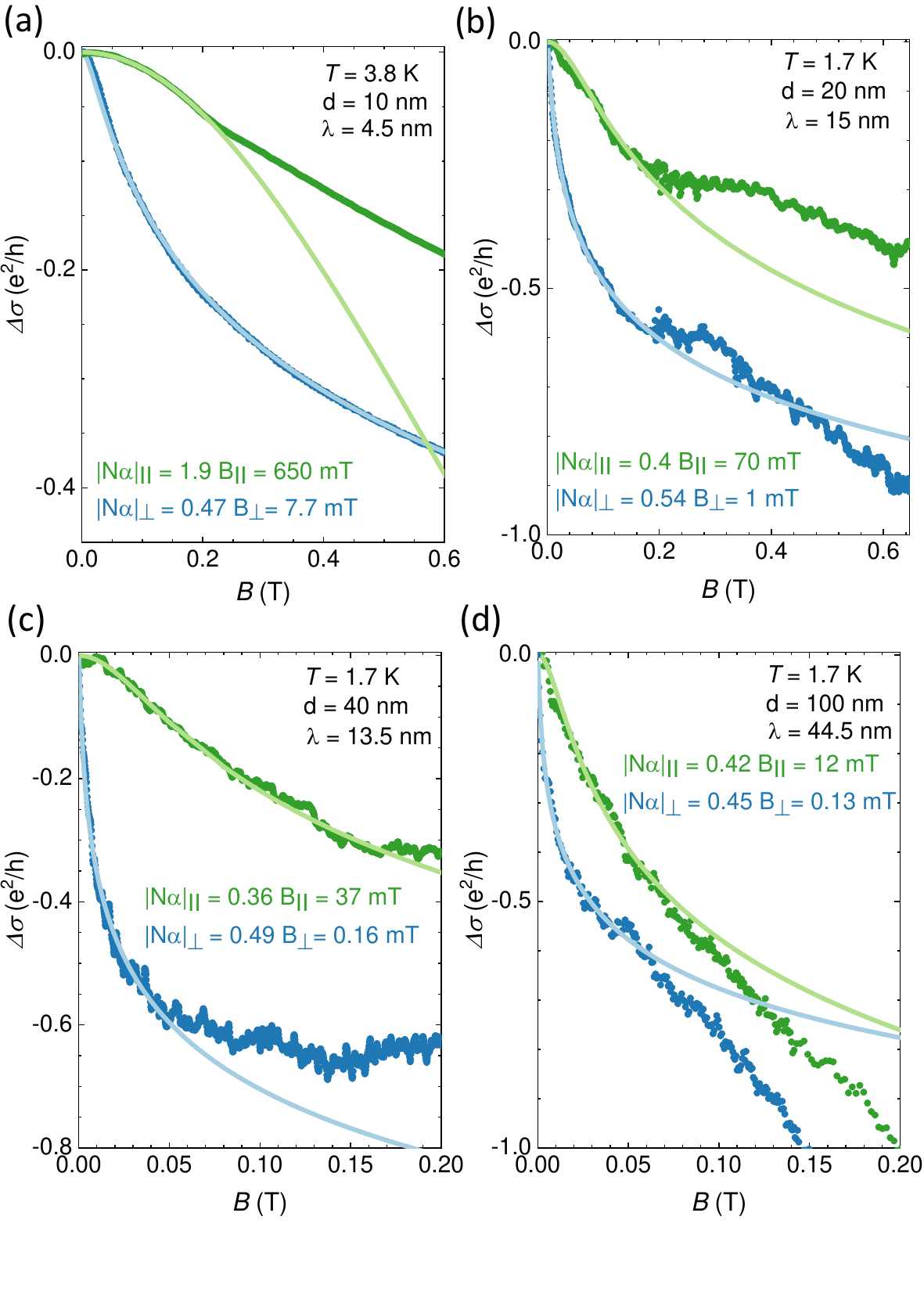}
	\caption{Panels (a)-(d) depict magnetoconductance for various sample thicknesses. Green points and lines present measurements for field parallel to the sample surface. Blue points and lines correspond to perpendicular field configuration. }
	\label{fig:PerpPara}
\end{figure}

The methodology of studying localization effects of the topological surface states  is similar to other two-dimensional systems. 
In particular, the ideal 2DEG is sensitive only to perpendicular component of the magnetic field. Thus, all the magnetoconductivity traces should merge when plotted as a function of $B\cos(\theta)$. The real observation is presented in Fig.~2(b), where region of coincidence of the data collected at different tilt angles is limited to very low magnetic fields. One of the possible cause was identified with finite width of 2DEG, which allows nonzero magnetic flux penetrating the 2DEG in the in-plane configuration. \cite{Altshuler1981,Dugaev1984,Beenakker1988,Raichev2000}  An additional ingredient was considered by  Sacksteder et al.,\cite{Sacksteder2014} who analyses the contribution of side walls which also host TSS's. Consequently, the tilted field dependence of magneto-conductivities should not coincide as a function of perpendicular field component.

\begin{figure}[tb]
	\includegraphics[width=\columnwidth]{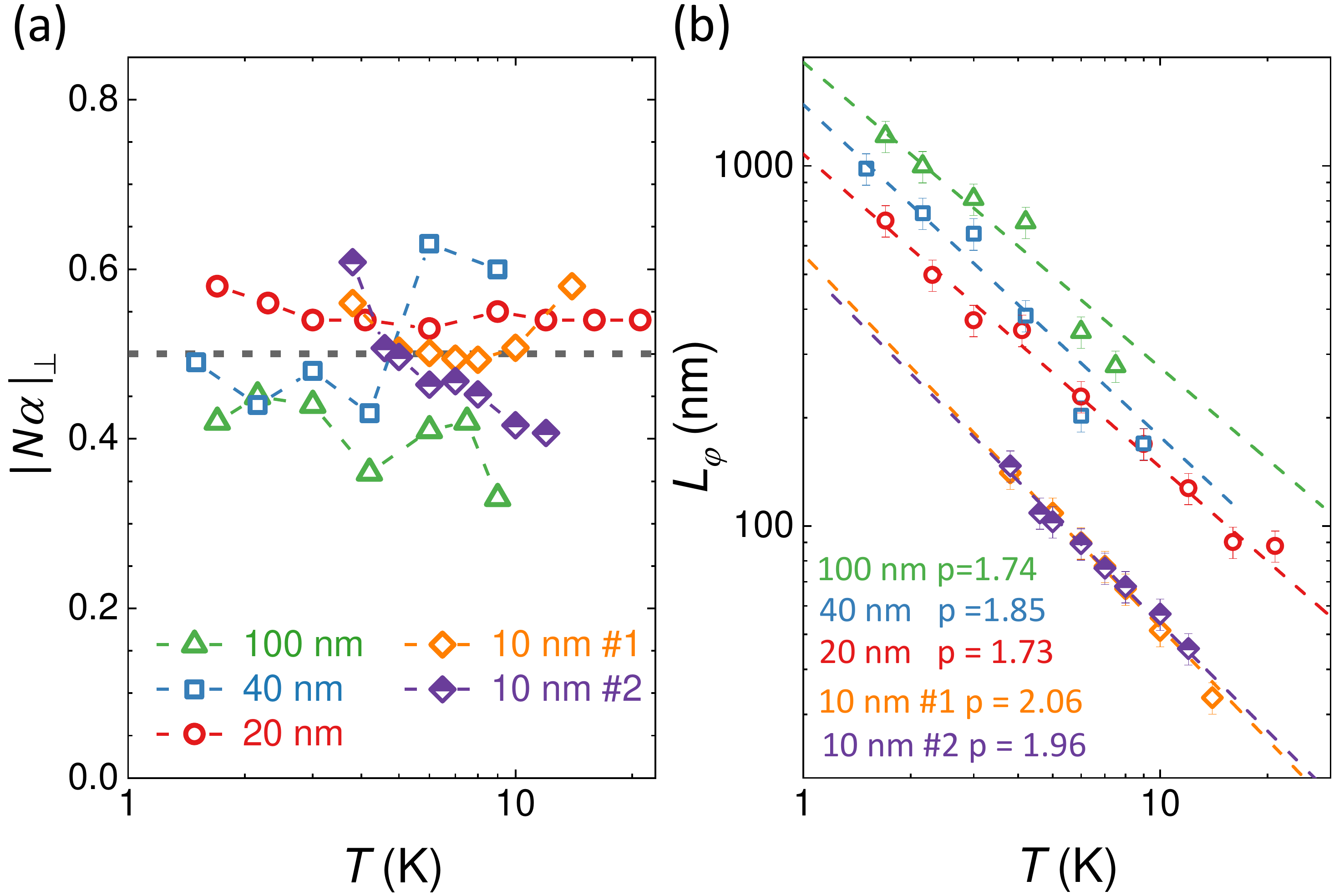}
	\caption{The results of fitting Eq.~\ref{OP} for different sample thickness at various temperatures (a) $|N\alpha|_{\perp}$ coefficient, grey dashed line corresponds to $|N\alpha|_{\perp}$=\,0.5,  (b) Phase coherence length $L_{\varphi}$. }
	\label{fig:Nalpha}
\end{figure}
The aforementioned papers use elastic mean free path ($L_{\rm tr}$) of the topological surface states in the description of WAL phenomenon. The conductivity of our samples has always a large contribution  from bulk states, therefore it is difficult to extract mean free path of TSS. To overcome this issue, we restrict data analysis to smallest number of fitting parameters. This is given by the model proposed by Tkachov and Hankiewicz \cite{Tkachov2011,Tkachov2013} of half-infinite slab with two-dimensional  TSS characterized by their decay length $\lambda$ into bulk. Additionaly, the model introduces only two more parameters -- phase coherence length $L_\varphi$ and $N\alpha$ - prefactor proportional to the number of independent coherent channels contributing to the conductivity. 
For completeness, we specify equations (\ref{OP}) and (\ref{IP}) as given in Ref. \onlinecite{Tkachov2013}:

\begin{equation}
\Delta\sigma_\perp (B)=N \alpha \frac{e^2}{\pi h}\Biggl[\psi\Biggl(\frac{1}{2} + \frac{ B_\perp }{B}\Biggr)-\ln\frac{ B_\perp }{B}\Biggr],\,B_\perp=\frac{ \hbar }{ 4|e| \, L^2_\varphi },
\label{OP}
\end{equation}
\begin{equation}
\Delta\sigma_\| (B) = N \alpha \frac{e^2}{\pi h}\ln\Biggl( 1 + \frac{B^2}{ B^2_{_\|} } \Biggr), \quad B_{_\|} = \frac{\hbar}{\sqrt{2} |e| \lambda L_\varphi }.
\label{IP}
\end{equation}
In both equations \ref{OP} and \ref{IP} the parameter $\alpha=-1/2$ denotes the symplectic universality class to which TSS belong.\cite{HLN1980}
It is multiplied by $N$ -- the number of independent channels contributing to total conductivity. The phase coherence length $L_\varphi$ is obtained directly from $B_\perp $ parameter. The parameter $\lambda$ introduced in Eq.~\ref{IP} is interpreted as decay length of the wave function of topological surface states. The explicit formula we quote below:

\begin{equation}
\lambda = \sqrt{\frac{2\hbar}{|e|}\frac{ B_\perp}{B^2_{_\|}}}.
 \label{lambda}
\end{equation}

Our fitting protocol relied on starting with very low magnetic field region and consecutively extending the fitted data region, simultaneously observing the values of parameters ($N\alpha$  and $B_{\perp}$ or $B_{\|}$). When parameters did not change much, the region of fit was further extended. The resulting theoretical lines are depicted in Fig.~2(c) and Fig.~3. We attribute the fit deviation at higher fields to the increasing contribution of 3D bulk states. 

\begin{figure}[t]
	\includegraphics[width=\columnwidth]{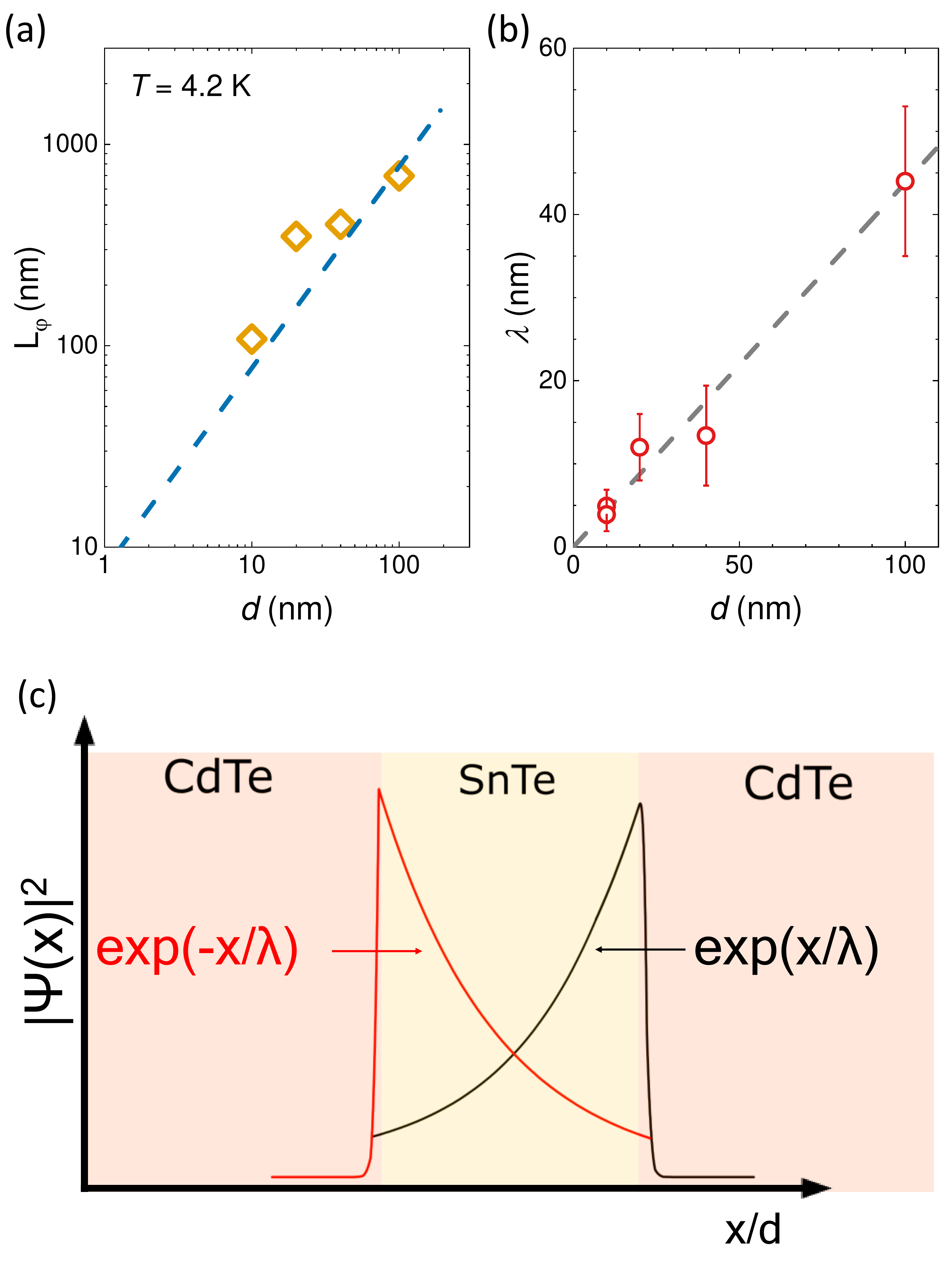}
	\caption{(a) Dephasing length extracted from the Eq.~\ref{OP} fit in the perpendicular configuration at 4.2\,K for various thicknesses. (b) $\lambda$ as calculated from Eq.~\ref{lambda} for as a function of film thickness. (c) Cartoon depicting modulus squared of the TSS wavefunction decaying exponentially into the bulk reservoir.}
	\label{fig:cartoon}
\end{figure}

In Fig. 4 we show the results of such  theoretical fits for perpendicular field configuration. In the situation studied, i.e.\,(001) oriented SnTe films,  we expected four conducting channels coming from four Dirac cones for both surfaces. Thus in the case of non-interacting channels the $|N \alpha |_{\perp}= 4 \times 2 \times 1/2 =4$. Instead we obtain with a good accuracy  $|N \alpha |_{\perp}=1/2$ for all samples studied in a broad range of temperatures (see Fig. 4(a)). 

We interpret this finding along scenario already pointed by Fukuyama \cite{Fukuyama1980} in early studies of Si-MOS magnetoconductivity. Namely, in the case  of scattering between $n_{v}$ valleys the coefficient  $n_v\alpha$ becomes renormalized to $n_v\alpha/n_v$ leaving in effect prefactor as for a single conducting channel $\alpha=-1/2$. 
The same argument applies to the situation of scattering between top and bottom TSS with the help of bulk states reservoir. \cite{Assaf2014,Dybko2017}

Fig.~4(b) shows the temperature dependence of the dephasing length fitted with exponential formula $L_{\varphi}\propto T^{-p/2}$. The exponent $p\simeq 2$ within experimental accuracy, which is characteristic  for electron--phonon interactions dominating at this temperature range.\cite{Belitz1987,Prinz1999}     
The existence of fully degenerate conduction channels is also confirmed in thickness dependence of dephasing length depicted in Fig. 5~(a). In case of non-interacting top and bottom TSS the dephasing length should be independent of thickness. Now in opposite case, the quantum interference path started at top TSS continues with the help of bulk reservoir states to the back TSS and returns to the starting point on top surface by scattering events within the bulk. This implies rising phase coherence length with sample thickness.

The main advantage of the Tkachov and Hankiewicz theory \cite{Tkachov2011,Tkachov2013} relies on supplying weak antilocalization formula (see Eq.~\ref{IP} ) for in-plane configuration. It enabled us to determine directly the decay length of TSS. The result of such an evaluation is summarized in Fig~5(b). Surprisingly we find $\lambda$ varying with the layer thickness. Without intersurface coupling one could expect $\lambda$ to be thickness independent. However, with our experimental accuracy we determine its value to be about 40\% of actual film thickness. The schematic picture is presented in Fig.~5(c), where square of the wave function of the TSS of both, top and bottom surfaces decay exponentially towards the bulk reservoir interior. We emphasize that TSS's wavefunctions do not hybridize with each other. The high values of $\lambda$ are consequence of interaction of TSS with the bulk reservoir.
The presented phenomenological description is in accord with recent tight binding calculations which describe wave function envelope of TSS on the border between trivial and inverted-band regions.\cite{Polley2018, Rechcinski2018}

{\it Conclusions.}
In summary, we report on growth of high quality SnTe (001) topological crystalline insulator layers on top of CdTe(4\,$\mu$m)//GaAs (001) substrate. The SnTe films were capped with 100\,nm thick CdTe layer to ensure the same trivial band -- TCI interface on both sides of the sample. The weak antilocalization measurements revealed the existence of the two-dimensional topological states.
We performed the analysis of low field magnetoconductivity in both perpendicular and in-plane magnetic field orientation. Within Tkachov and Hankiewicz model we found decoherence length depending on the SnTe layer thickness and only one coherent channel taking part in conduction instead of eight expected. Our interpretation relies on  assumption of intervalley scattering within one topological surface and intersurface scattering mediated by bulk reservoir. Both lead to eight-fold degeneracy which renormalizes the coefficient in WAL expression to one characteristic for single channel only. In this communication we extended the WAL data analysis by additional parameter describing the spatial extent of the wave function of TSS. We found it depending linearly on the SnTe layer thickness, presumably on account of scattering with bulk reservoir.     
We anticipate that for thinner SnTe layers the spatial extent of the TSS will reach  hybrydization limit for top and bottom surfaces, which should be observed as weak antilocalization to weak localization crossover.\cite{Linder2009,HaiZhouLu2010,Zhang2010} 

{\it Acknowledgments.}
We thank Ryszard Buczko, Maciej Sawicki and Tomasz Dietl for helpful discussions.
The research was partially supported by the Foundation for Polish Science through the IRA Programme co-financed by EU within SG OP. The work at the Institute of Physics, Polish Academy of Sciences was supported by the National Science Center (Poland) through the grants: PRELUDIUM (2015/19/N/ST3/02626), OPUS (2014/15/B/ST3/03833, 2017/27/B/ST3/02470).


\begin{thebibliography}{34}%
	\makeatletter
	\providecommand \@ifxundefined [1]{%
		\@ifx{#1\undefined}
	}%
	\providecommand \@ifnum [1]{%
		\ifnum #1\expandafter \@firstoftwo
		\else \expandafter \@secondoftwo
		\fi
	}%
	\providecommand \@ifx [1]{%
		\ifx #1\expandafter \@firstoftwo
		\else \expandafter \@secondoftwo
		\fi
	}%
	\providecommand \natexlab [1]{#1}%
	\providecommand \enquote  [1]{``#1''}%
	\providecommand \bibnamefont  [1]{#1}%
	\providecommand \bibfnamefont [1]{#1}%
	\providecommand \citenamefont [1]{#1}%
	\providecommand \href@noop [0]{\@secondoftwo}%
	\providecommand \href [0]{\begingroup \@sanitize@url \@href}%
	\providecommand \@href[1]{\@@startlink{#1}\@@href}%
	\providecommand \@@href[1]{\endgroup#1\@@endlink}%
	\providecommand \@sanitize@url [0]{\catcode `\\12\catcode `\$12\catcode
		`\&12\catcode `\#12\catcode `\^12\catcode `\_12\catcode `\%12\relax}%
	\providecommand \@@startlink[1]{}%
	\providecommand \@@endlink[0]{}%
	\providecommand \url  [0]{\begingroup\@sanitize@url \@url }%
	\providecommand \@url [1]{\endgroup\@href {#1}{\urlprefix }}%
	\providecommand \urlprefix  [0]{URL }%
	\providecommand \Eprint [0]{\href }%
	\providecommand \doibase [0]{http://dx.doi.org/}%
	\providecommand \selectlanguage [0]{\@gobble}%
	\providecommand \bibinfo  [0]{\@secondoftwo}%
	\providecommand \bibfield  [0]{\@secondoftwo}%
	\providecommand \translation [1]{[#1]}%
	\providecommand \BibitemOpen [0]{}%
	\providecommand \bibitemStop [0]{}%
	\providecommand \bibitemNoStop [0]{.\EOS\space}%
	\providecommand \EOS [0]{\spacefactor3000\relax}%
	\providecommand \BibitemShut  [1]{\csname bibitem#1\endcsname}%
	\let\auto@bib@innerbib\@empty
	\bibitem [{Mol(2013)}]{Molen2013i}%
	\BibitemOpen
	\bibfield  {title} {\enquote {\bibinfo {title} {Topological insulators},}\
	}in\ \href {\doibase https://doi.org/10.1016/B978-0-444-63314-9.00012-3}
	{\emph {\bibinfo {booktitle} {Topological Insulators}}},\ \bibinfo {series}
	{Contemporary Concepts of Condensed Matter Science}, Vol.~\bibinfo {volume}
	{6},\ \bibinfo {editor} {edited by\ \bibinfo {editor} {\bibfnamefont
			{Marcel}\ \bibnamefont {Franz}}\ and\ \bibinfo {editor} {\bibfnamefont
			{Laurens}\ \bibnamefont {Molenkamp}}}\ (\bibinfo  {publisher} {Elsevier},\
	\bibinfo {year} {2013})\BibitemShut {NoStop}%
	\bibitem [{\citenamefont {Ando}\ and\ \citenamefont {Fu}(2015)}]{Ando2015}%
	\BibitemOpen
	\bibfield  {author} {\bibinfo {author} {\bibfnamefont {Yoichi}\ \bibnamefont
			{Ando}}\ and\ \bibinfo {author} {\bibfnamefont {Liang}\ \bibnamefont {Fu}},\
	}\bibfield  {title} {\enquote {\bibinfo {title} {{Topological Crystalline
					Insulators and Topological Superconductors: From Concepts to Materials}},}\
	}\href {\doibase 10.1146/annurev-conmatphys-031214-014501} {\bibfield
		{journal} {\bibinfo  {journal} {Annu. Rev. Condens. Matter Phys.}\ }\textbf
		{\bibinfo {volume} {6}},\ \bibinfo {pages} {361--381} (\bibinfo {year}
		{2015})}\BibitemShut {NoStop}%
	\bibitem [{\citenamefont {Fu}(2011)}]{Fu2010}%
	\BibitemOpen
	\bibfield  {author} {\bibinfo {author} {\bibfnamefont {Liang}\ \bibnamefont
			{Fu}},\ }\bibfield  {title} {\enquote {\bibinfo {title} {Topological
				crystalline insulators},}\ }\href {\doibase 10.1103/PhysRevLett.106.106802}
	{\bibfield  {journal} {\bibinfo  {journal} {Phys. Rev. Lett.}\ }\textbf
		{\bibinfo {volume} {106}},\ \bibinfo {pages} {106802} (\bibinfo {year}
		{2011})}\BibitemShut {NoStop}%
	\bibitem [{\citenamefont {Hsieh}\ \emph {et~al.}(2012)\citenamefont {Hsieh},
		\citenamefont {Lin}, \citenamefont {Liu}, \citenamefont {Duan}, \citenamefont
		{Bansil},\ and\ \citenamefont {Fu}}]{Hsieh2012}%
	\BibitemOpen
	\bibfield  {author} {\bibinfo {author} {\bibfnamefont {Timothy~H}\
			\bibnamefont {Hsieh}}, \bibinfo {author} {\bibfnamefont {Hsin}\ \bibnamefont
			{Lin}}, \bibinfo {author} {\bibfnamefont {Junwei}\ \bibnamefont {Liu}},
		\bibinfo {author} {\bibfnamefont {Wenhui}\ \bibnamefont {Duan}}, \bibinfo
		{author} {\bibfnamefont {Arun}\ \bibnamefont {Bansil}}, \ and\ \bibinfo
		{author} {\bibfnamefont {Liang}\ \bibnamefont {Fu}},\ }\bibfield  {title}
	{\enquote {\bibinfo {title} {{Topological crystalline insulators in the SnTe
					material class.}}}\ }\href {\doibase 10.1038/ncomms1969} {\bibfield
		{journal} {\bibinfo  {journal} {Nat. Commun.}\ }\textbf {\bibinfo {volume}
			{3}},\ \bibinfo {pages} {982} (\bibinfo {year} {2012})}\BibitemShut {NoStop}%
	\bibitem [{\citenamefont {Dziawa}\ \emph {et~al.}(2012)\citenamefont {Dziawa},
		\citenamefont {Kowalski}, \citenamefont {Dybko}, \citenamefont {Buczko},
		\citenamefont {Szczerbakow}, \citenamefont {Szot}, \citenamefont
		{{\L}usakowska}, \citenamefont {Balasubramanian}, \citenamefont {Wojek},
		\citenamefont {Berntsen}, \citenamefont {Tjernberg},\ and\ \citenamefont
		{Story}}]{Dziawa2012}%
	\BibitemOpen
	\bibfield  {author} {\bibinfo {author} {\bibfnamefont {P.}~\bibnamefont
			{Dziawa}}, \bibinfo {author} {\bibfnamefont {B.~J.}\ \bibnamefont
			{Kowalski}}, \bibinfo {author} {\bibfnamefont {K.}~\bibnamefont {Dybko}},
		\bibinfo {author} {\bibfnamefont {R.}~\bibnamefont {Buczko}}, \bibinfo
		{author} {\bibfnamefont {A.}~\bibnamefont {Szczerbakow}}, \bibinfo {author}
		{\bibfnamefont {M.}~\bibnamefont {Szot}}, \bibinfo {author} {\bibfnamefont
			{E.}~\bibnamefont {{\L}usakowska}}, \bibinfo {author} {\bibfnamefont
			{T.}~\bibnamefont {Balasubramanian}}, \bibinfo {author} {\bibfnamefont
			{B.~M.}\ \bibnamefont {Wojek}}, \bibinfo {author} {\bibfnamefont {M.~H.}\
			\bibnamefont {Berntsen}}, \bibinfo {author} {\bibfnamefont {O.}~\bibnamefont
			{Tjernberg}}, \ and\ \bibinfo {author} {\bibfnamefont {T.}~\bibnamefont
			{Story}},\ }\bibfield  {title} {\enquote {\bibinfo {title} {Topological
				crystalline insulator states in {Pb$_{1-x}$Sn$_x$Se}},}\ }\href {\doibase
		10.1038/nmat3449} {\bibfield  {journal} {\bibinfo  {journal} {Nat. Mater.}\
		}\textbf {\bibinfo {volume} {11}},\ \bibinfo {pages} {1023--1027} (\bibinfo
		{year} {2012})}\BibitemShut {NoStop}%
	\bibitem [{\citenamefont {Xu}\ \emph {et~al.}(2012)\citenamefont {Xu},
		\citenamefont {Liu}, \citenamefont {Alidoust}, \citenamefont {Neupane},
		\citenamefont {Qian}, \citenamefont {Belopolski}, \citenamefont {Denlinger},
		\citenamefont {Wang}, \citenamefont {H.~Lin}, \citenamefont {Landolt},
		\citenamefont {Slomski}, \citenamefont {Dil}, \citenamefont {Marcinkova},
		\citenamefont {Morosan}, \citenamefont {Gibson}, \citenamefont {Sankar},
		\citenamefont {Chou}, \citenamefont {Cava}, \citenamefont {Bansil},\ and\
		\citenamefont {Hasan}}]{Xu2012}%
	\BibitemOpen
	\bibfield  {author} {\bibinfo {author} {\bibfnamefont {Su-Yang}\ \bibnamefont
			{Xu}}, \bibinfo {author} {\bibfnamefont {Chang}\ \bibnamefont {Liu}},
		\bibinfo {author} {\bibfnamefont {N.}~\bibnamefont {Alidoust}}, \bibinfo
		{author} {\bibfnamefont {M.}~\bibnamefont {Neupane}}, \bibinfo {author}
		{\bibfnamefont {D.}~\bibnamefont {Qian}}, \bibinfo {author} {\bibfnamefont
			{I.}~\bibnamefont {Belopolski}}, \bibinfo {author} {\bibfnamefont {J.D.}\
			\bibnamefont {Denlinger}}, \bibinfo {author} {\bibfnamefont {Y.J.}\
			\bibnamefont {Wang}}, \bibinfo {author} {\bibfnamefont {L.A.~Wray}\
			\bibnamefont {H.~Lin}}, \bibinfo {author} {\bibfnamefont {G.}~\bibnamefont
			{Landolt}}, \bibinfo {author} {\bibfnamefont {B.}~\bibnamefont {Slomski}},
		\bibinfo {author} {\bibfnamefont {J.H.}\ \bibnamefont {Dil}}, \bibinfo
		{author} {\bibfnamefont {A.}~\bibnamefont {Marcinkova}}, \bibinfo {author}
		{\bibfnamefont {E.}~\bibnamefont {Morosan}}, \bibinfo {author} {\bibfnamefont
			{Q.}~\bibnamefont {Gibson}}, \bibinfo {author} {\bibfnamefont
			{R.}~\bibnamefont {Sankar}}, \bibinfo {author} {\bibfnamefont {F.C.}\
			\bibnamefont {Chou}}, \bibinfo {author} {\bibfnamefont {R.J.}\ \bibnamefont
			{Cava}}, \bibinfo {author} {\bibfnamefont {A.}~\bibnamefont {Bansil}}, \ and\
		\bibinfo {author} {\bibfnamefont {M.Z.}\ \bibnamefont {Hasan}},\ }\bibfield
	{title} {\enquote {\bibinfo {title} {Observation of a topological crystalline
				insulator phase and topological phase transition in {Pb$_{1-x}$Sn$_x$Te}},}\
	}\href {\doibase 10.1038/ncomms2191} {\bibfield  {journal} {\bibinfo
			{journal} {Nat. Commun.}\ }\textbf {\bibinfo {volume} {3}},\ \bibinfo {pages}
		{1192} (\bibinfo {year} {2012})}\BibitemShut {NoStop}%
	\bibitem [{\citenamefont {Tanaka}\ \emph {et~al.}(2012)\citenamefont {Tanaka},
		\citenamefont {Ren}, \citenamefont {Sato}, \citenamefont {Nakayama},
		\citenamefont {Souma}, \citenamefont {Takahashi}, \citenamefont {Segawa},\
		and\ \citenamefont {Ando}}]{Tanaka2012}%
	\BibitemOpen
	\bibfield  {author} {\bibinfo {author} {\bibfnamefont {Y.}~\bibnamefont
			{Tanaka}}, \bibinfo {author} {\bibfnamefont {Zhi}\ \bibnamefont {Ren}},
		\bibinfo {author} {\bibfnamefont {T.}~\bibnamefont {Sato}}, \bibinfo {author}
		{\bibfnamefont {K.}~\bibnamefont {Nakayama}}, \bibinfo {author}
		{\bibfnamefont {S.}~\bibnamefont {Souma}}, \bibinfo {author} {\bibfnamefont
			{T.}~\bibnamefont {Takahashi}}, \bibinfo {author} {\bibfnamefont {Kouji}\
			\bibnamefont {Segawa}}, \ and\ \bibinfo {author} {\bibfnamefont {Yoichi}\
			\bibnamefont {Ando}},\ }\bibfield  {title} {\enquote {\bibinfo {title}
			{{Experimental realization of a topological crystalline insulator in
					SnTe}},}\ }\href {\doibase 10.1038/nphys2442} {\bibfield  {journal} {\bibinfo
			{journal} {Nat. Phys.}\ }\textbf {\bibinfo {volume} {8}},\ \bibinfo {pages}
		{800--803} (\bibinfo {year} {2012})}\BibitemShut {NoStop}%
	\bibitem [{\citenamefont {Hsieh}\ \emph {et~al.}(2009)\citenamefont {Hsieh},
		\citenamefont {Xia}, \citenamefont {Wray}, \citenamefont {Qian},
		\citenamefont {Pal}, \citenamefont {Dil}, \citenamefont {Osterwalder},
		\citenamefont {Meier}, \citenamefont {Bihlmayer}, \citenamefont {Kane},
		\citenamefont {Hor}, \citenamefont {Cava},\ and\ \citenamefont
		{Hasan}}]{Hsieh2009}%
	\BibitemOpen
	\bibfield  {author} {\bibinfo {author} {\bibfnamefont {D.}~\bibnamefont
			{Hsieh}}, \bibinfo {author} {\bibfnamefont {Y.}~\bibnamefont {Xia}}, \bibinfo
		{author} {\bibfnamefont {L.}~\bibnamefont {Wray}}, \bibinfo {author}
		{\bibfnamefont {D.}~\bibnamefont {Qian}}, \bibinfo {author} {\bibfnamefont
			{A.}~\bibnamefont {Pal}}, \bibinfo {author} {\bibfnamefont {J.~H.}\
			\bibnamefont {Dil}}, \bibinfo {author} {\bibfnamefont {J.}~\bibnamefont
			{Osterwalder}}, \bibinfo {author} {\bibfnamefont {F.}~\bibnamefont {Meier}},
		\bibinfo {author} {\bibfnamefont {G.}~\bibnamefont {Bihlmayer}}, \bibinfo
		{author} {\bibfnamefont {C.~L.}\ \bibnamefont {Kane}}, \bibinfo {author}
		{\bibfnamefont {Y.~S.}\ \bibnamefont {Hor}}, \bibinfo {author} {\bibfnamefont
			{R.~J.}\ \bibnamefont {Cava}}, \ and\ \bibinfo {author} {\bibfnamefont
			{M.~Z.}\ \bibnamefont {Hasan}},\ }\bibfield  {title} {\enquote {\bibinfo
			{title} {Observation of unconventional quantum spin textures in topological
				insulators},}\ }\href {\doibase 10.1126/science.1167733} {\bibfield
		{journal} {\bibinfo  {journal} {Science}\ }\textbf {\bibinfo {volume}
			{323}},\ \bibinfo {pages} {919--922} (\bibinfo {year} {2009})}\BibitemShut
	{NoStop}%
	\bibitem [{\citenamefont {Dybko}\ \emph {et~al.}(2017)\citenamefont {Dybko},
		\citenamefont {Szot}, \citenamefont {Szczerbakow}, \citenamefont {Gutowska},
		\citenamefont {Zajarniuk}, \citenamefont {Domagala}, \citenamefont
		{Szewczyk}, \citenamefont {Story},\ and\ \citenamefont
		{Zawadzki}}]{Dybko2017}%
	\BibitemOpen
	\bibfield  {author} {\bibinfo {author} {\bibfnamefont {K.}~\bibnamefont
			{Dybko}}, \bibinfo {author} {\bibfnamefont {M.}~\bibnamefont {Szot}},
		\bibinfo {author} {\bibfnamefont {A.}~\bibnamefont {Szczerbakow}}, \bibinfo
		{author} {\bibfnamefont {M.~U.}\ \bibnamefont {Gutowska}}, \bibinfo {author}
		{\bibfnamefont {T.}~\bibnamefont {Zajarniuk}}, \bibinfo {author}
		{\bibfnamefont {J.~Z.}\ \bibnamefont {Domagala}}, \bibinfo {author}
		{\bibfnamefont {A.}~\bibnamefont {Szewczyk}}, \bibinfo {author}
		{\bibfnamefont {T.}~\bibnamefont {Story}}, \ and\ \bibinfo {author}
		{\bibfnamefont {W.}~\bibnamefont {Zawadzki}},\ }\bibfield  {title} {\enquote
		{\bibinfo {title} {Experimental evidence for topological surface states
				wrapping around a bulk {SnTe} crystal},}\ }\href {\doibase
		10.1103/PhysRevB.96.205129} {\bibfield  {journal} {\bibinfo  {journal} {Phys.
				Rev. B}\ }\textbf {\bibinfo {volume} {96}},\ \bibinfo {pages} {205129}
		(\bibinfo {year} {2017})}\BibitemShut {NoStop}%
	\bibitem [{\citenamefont {Iordanskii}\ \emph {et~al.}(1994)\citenamefont
		{Iordanskii}, \citenamefont {Lyanda-Geller},\ and\ \citenamefont
		{Pikus}}]{iordanskii1994}%
	\BibitemOpen
	\bibfield  {author} {\bibinfo {author} {\bibfnamefont {S.V.}\ \bibnamefont
			{Iordanskii}}, \bibinfo {author} {\bibfnamefont {Y.~B.}\ \bibnamefont
			{Lyanda-Geller}}, \ and\ \bibinfo {author} {\bibfnamefont {G.E.}\
			\bibnamefont {Pikus}},\ }\bibfield  {title} {\enquote {\bibinfo {title} {Weak
				localization in quantum wells with spin-orbit interaction},}\ }\href@noop {}
	{\bibfield  {journal} {\bibinfo  {journal} {JETP Letters}\ }\textbf {\bibinfo
			{volume} {60}},\ \bibinfo {pages} {199} (\bibinfo {year} {1994})}\BibitemShut
	{NoStop}%
	\bibitem [{\citenamefont {Hikami}\ \emph {et~al.}(1980)\citenamefont {Hikami},
		\citenamefont {Larkin},\ and\ \citenamefont {Nagaoka}}]{HLN1980}%
	\BibitemOpen
	\bibfield  {author} {\bibinfo {author} {\bibfnamefont {S.}~\bibnamefont
			{Hikami}}, \bibinfo {author} {\bibfnamefont {A.~I.}\ \bibnamefont {Larkin}},
		\ and\ \bibinfo {author} {\bibfnamefont {Y.}~\bibnamefont {Nagaoka}},\
	}\bibfield  {title} {\enquote {\bibinfo {title} {Spin-orbit interaction and
				magnetoresistance in the two dimensional random system},}\ }\href {\doibase
		10.1143/PTP.63.707} {\bibfield  {journal} {\bibinfo  {journal} {Progress of
				Theoretical Physics}\ }\textbf {\bibinfo {volume} {63}},\ \bibinfo {pages}
		{707--710} (\bibinfo {year} {1980})}\BibitemShut {NoStop}%
	\bibitem [{\citenamefont {Kim}\ \emph {et~al.}(2011)\citenamefont {Kim},
		\citenamefont {Brahlek}, \citenamefont {Bansal}, \citenamefont {Edrey},
		\citenamefont {Kapilevich}, \citenamefont {Iida}, \citenamefont {Tanimura},
		\citenamefont {Horibe}, \citenamefont {Cheong},\ and\ \citenamefont
		{Oh}}]{Kim2011}%
	\BibitemOpen
	\bibfield  {author} {\bibinfo {author} {\bibfnamefont {Yong~Seung}\
			\bibnamefont {Kim}}, \bibinfo {author} {\bibfnamefont {M}~\bibnamefont
			{Brahlek}}, \bibinfo {author} {\bibfnamefont {N}~\bibnamefont {Bansal}},
		\bibinfo {author} {\bibfnamefont {Eliav}\ \bibnamefont {Edrey}}, \bibinfo
		{author} {\bibfnamefont {Gary~A.}\ \bibnamefont {Kapilevich}}, \bibinfo
		{author} {\bibfnamefont {Keiko}\ \bibnamefont {Iida}}, \bibinfo {author}
		{\bibfnamefont {Makoto}\ \bibnamefont {Tanimura}}, \bibinfo {author}
		{\bibfnamefont {Yoichi}\ \bibnamefont {Horibe}}, \bibinfo {author}
		{\bibfnamefont {Sang-Wook}\ \bibnamefont {Cheong}}, \ and\ \bibinfo {author}
		{\bibfnamefont {Seongshik}\ \bibnamefont {Oh}},\ }\bibfield  {title}
	{\enquote {\bibinfo {title} {Thickness-dependent bulk properties and weak
				antilocalization effect in topological insulator {Bi${}_{2}$Se${}_{3}$}},}\
	}\href {\doibase 10.1103/PhysRevB.84.073109} {\bibfield  {journal} {\bibinfo
			{journal} {Phys. Rev. B}\ }\textbf {\bibinfo {volume} {84}},\ \bibinfo
		{pages} {073109} (\bibinfo {year} {2011})}\BibitemShut {NoStop}%
	\bibitem [{\citenamefont {Lin}\ \emph {et~al.}(2013)\citenamefont {Lin},
		\citenamefont {He}, \citenamefont {Liao}, \citenamefont {Wang}, \citenamefont
		{IV}, \citenamefont {Yang}, \citenamefont {Guan}, \citenamefont {Zhang},
		\citenamefont {Gu}, \citenamefont {Zhang}, \citenamefont {Zeng},
		\citenamefont {Dai}, \citenamefont {Wu},\ and\ \citenamefont
		{Li}}]{Sacksteder2013}%
	\BibitemOpen
	\bibfield  {author} {\bibinfo {author} {\bibfnamefont {C.~J.}\ \bibnamefont
			{Lin}}, \bibinfo {author} {\bibfnamefont {X.~Y.}\ \bibnamefont {He}},
		\bibinfo {author} {\bibfnamefont {J.}~\bibnamefont {Liao}}, \bibinfo {author}
		{\bibfnamefont {X.~X.}\ \bibnamefont {Wang}}, \bibinfo {author}
		{\bibfnamefont {V.~Sacksteder}\ \bibnamefont {IV}}, \bibinfo {author}
		{\bibfnamefont {W.~M.}\ \bibnamefont {Yang}}, \bibinfo {author}
		{\bibfnamefont {T.}~\bibnamefont {Guan}}, \bibinfo {author} {\bibfnamefont
			{Q.~M.}\ \bibnamefont {Zhang}}, \bibinfo {author} {\bibfnamefont
			{L.}~\bibnamefont {Gu}}, \bibinfo {author} {\bibfnamefont {G.~Y.}\
			\bibnamefont {Zhang}}, \bibinfo {author} {\bibfnamefont {C.~G.}\ \bibnamefont
			{Zeng}}, \bibinfo {author} {\bibfnamefont {X.}~\bibnamefont {Dai}}, \bibinfo
		{author} {\bibfnamefont {K.~H.}\ \bibnamefont {Wu}}, \ and\ \bibinfo {author}
		{\bibfnamefont {Y.~Q.}\ \bibnamefont {Li}},\ }\bibfield  {title} {\enquote
		{\bibinfo {title} {Parallel field magnetoresistance in topological insulator
				thin films},}\ }\href {\doibase 10.1103/PhysRevB.88.041307} {\bibfield
		{journal} {\bibinfo  {journal} {Phys. Rev. B}\ }\textbf {\bibinfo {volume}
			{88}},\ \bibinfo {pages} {041307} (\bibinfo {year} {2013})}\BibitemShut
	{NoStop}%
	\bibitem [{\citenamefont {Assaf}\ \emph {et~al.}(2014)\citenamefont {Assaf},
		\citenamefont {Katmis}, \citenamefont {Wei}, \citenamefont {Satpati},
		\citenamefont {Zhang}, \citenamefont {Bennett}, \citenamefont {Harris},
		\citenamefont {Moodera},\ and\ \citenamefont {Heiman}}]{Assaf2014}%
	\BibitemOpen
	\bibfield  {author} {\bibinfo {author} {\bibfnamefont {B.~A.}\ \bibnamefont
			{Assaf}}, \bibinfo {author} {\bibfnamefont {F.}~\bibnamefont {Katmis}},
		\bibinfo {author} {\bibfnamefont {P.}~\bibnamefont {Wei}}, \bibinfo {author}
		{\bibfnamefont {B.}~\bibnamefont {Satpati}}, \bibinfo {author} {\bibfnamefont
			{Z.}~\bibnamefont {Zhang}}, \bibinfo {author} {\bibfnamefont {S.~P.}\
			\bibnamefont {Bennett}}, \bibinfo {author} {\bibfnamefont {V.~G.}\
			\bibnamefont {Harris}}, \bibinfo {author} {\bibfnamefont {J.~S.}\
			\bibnamefont {Moodera}}, \ and\ \bibinfo {author} {\bibfnamefont
			{D.}~\bibnamefont {Heiman}},\ }\bibfield  {title} {\enquote {\bibinfo {title}
			{{Quantum coherent transport in SnTe topological crystalline insulator thin
					films}},}\ }\href {\doibase 10.1063/1.4895456} {\bibfield  {journal}
		{\bibinfo  {journal} {Appl. Phys. Lett.}\ }\textbf {\bibinfo {volume}
			{105}},\ \bibinfo {pages} {102108} (\bibinfo {year} {2014})}\BibitemShut
	{NoStop}%
	\bibitem [{\citenamefont {Akiyama}\ \emph {et~al.}(2016)\citenamefont
		{Akiyama}, \citenamefont {Fujisawa}, \citenamefont {Yamaguchi}, \citenamefont
		{Ishikawa},\ and\ \citenamefont {Kuroda}}]{Akiyama2016}%
	\BibitemOpen
	\bibfield  {author} {\bibinfo {author} {\bibfnamefont {R.}~\bibnamefont
			{Akiyama}}, \bibinfo {author} {\bibfnamefont {K.}~\bibnamefont {Fujisawa}},
		\bibinfo {author} {\bibfnamefont {T.}~\bibnamefont {Yamaguchi}}, \bibinfo
		{author} {\bibfnamefont {R.}~\bibnamefont {Ishikawa}}, \ and\ \bibinfo
		{author} {\bibfnamefont {S.}~\bibnamefont {Kuroda}},\ }\bibfield  {title}
	{\enquote {\bibinfo {title} {{Two-dimensional quantum transport of
					multivalley (111) surface state in topological crystalline insulator SnTe
					thin films}},}\ }\href {\doibase 10.1007/s12274-015-0930-8} {\bibfield
		{journal} {\bibinfo  {journal} {Nano Res.}\ }\textbf {\bibinfo {volume}
			{9}},\ \bibinfo {pages} {490--498} (\bibinfo {year} {2016})}\BibitemShut
	{NoStop}%
	\bibitem [{\citenamefont {Ishikawa}\ \emph {et~al.}(2016)\citenamefont
		{Ishikawa}, \citenamefont {Yamaguchi}, \citenamefont {Ohtaki}, \citenamefont
		{Akiyama},\ and\ \citenamefont {Kuroda}}]{Ishikawa2016}%
	\BibitemOpen
	\bibfield  {author} {\bibinfo {author} {\bibfnamefont {R.}~\bibnamefont
			{Ishikawa}}, \bibinfo {author} {\bibfnamefont {T.}~\bibnamefont {Yamaguchi}},
		\bibinfo {author} {\bibfnamefont {Y.}~\bibnamefont {Ohtaki}}, \bibinfo
		{author} {\bibfnamefont {R.}~\bibnamefont {Akiyama}}, \ and\ \bibinfo
		{author} {\bibfnamefont {S.}~\bibnamefont {Kuroda}},\ }\bibfield  {title}
	{\enquote {\bibinfo {title} {Thin film growth of a topological crystal
				insulator {SnTe} on the {CdTe} (111) surface by molecular beam epitaxy},}\
	}\href {\doibase https://doi.org/10.1016/j.jcrysgro.2016.08.027} {\bibfield
		{journal} {\bibinfo  {journal} {Journal of Crystal Growth}\ }\textbf
		{\bibinfo {volume} {453}},\ \bibinfo {pages} {124 -- 129} (\bibinfo {year}
		{2016})}\BibitemShut {NoStop}%
	\bibitem [{\citenamefont {Tkachov}\ and\ \citenamefont
		{Hankiewicz}(2011)}]{Tkachov2011}%
	\BibitemOpen
	\bibfield  {author} {\bibinfo {author} {\bibfnamefont {G.}~\bibnamefont
			{Tkachov}}\ and\ \bibinfo {author} {\bibfnamefont {E.~M.}\ \bibnamefont
			{Hankiewicz}},\ }\bibfield  {title} {\enquote {\bibinfo {title} {{Weak
					antilocalization in HgTe quantum wells and topological surface states:
					Massive versus massless Dirac fermions}},}\ }\href {\doibase
		10.1103/PhysRevB.84.035444} {\bibfield  {journal} {\bibinfo  {journal} {Phys.
				Rev. B}\ }\textbf {\bibinfo {volume} {84}},\ \bibinfo {pages} {035444}
		(\bibinfo {year} {2011})}\BibitemShut {NoStop}%
	\bibitem [{\citenamefont {Tkachov}\ and\ \citenamefont
		{Hankiewicz}(2013)}]{Tkachov2013}%
	\BibitemOpen
	\bibfield  {author} {\bibinfo {author} {\bibfnamefont {G.}~\bibnamefont
			{Tkachov}}\ and\ \bibinfo {author} {\bibfnamefont {E.~M.}\ \bibnamefont
			{Hankiewicz}},\ }\bibfield  {title} {\enquote {\bibinfo {title}
			{{Spin-helical transport in normal and superconducting topological
					insulators}},}\ }\href {\doibase 10.1002/pssb.201248385} {\bibfield
		{journal} {\bibinfo  {journal} {Phys. Status Solidi}\ }\textbf {\bibinfo
			{volume} {250}},\ \bibinfo {pages} {215--232} (\bibinfo {year}
		{2013})}\BibitemShut {NoStop}%
	\bibitem [{\citenamefont {Maekawa}\ and\ \citenamefont
		{Fukuyama}(1982)}]{Maekawa1981}%
	\BibitemOpen
	\bibfield  {author} {\bibinfo {author} {\bibfnamefont {Sadamishi}\
			\bibnamefont {Maekawa}}\ and\ \bibinfo {author} {\bibfnamefont {Hidetoshi}\
			\bibnamefont {Fukuyama}},\ }\bibfield  {title} {\enquote {\bibinfo {title}
			{Localization effects in two-dimensional superconductors},}\ }\href {\doibase
		10.1143/JPSJ.51.1380} {\bibfield  {journal} {\bibinfo  {journal} {Journal of
				the Physical Society of Japan}\ }\textbf {\bibinfo {volume} {51}},\ \bibinfo
		{pages} {1380--1385} (\bibinfo {year} {1982})}\BibitemShut {NoStop}%
	\bibitem [{\citenamefont {Altshuler}\ and\ \citenamefont
		{Aronov}(1981)}]{Altshuler1981}%
	\BibitemOpen
	\bibfield  {author} {\bibinfo {author} {\bibfnamefont {B.~L.}\ \bibnamefont
			{Altshuler}}\ and\ \bibinfo {author} {\bibfnamefont {A.~G.}\ \bibnamefont
			{Aronov}},\ }\bibfield  {title} {\enquote {\bibinfo {title}
			{{Magnetoresistance of thin films and of wires in a longitudinal magnetic
					field}},}\ }\href@noop {} {\bibfield  {journal} {\bibinfo  {journal} {JETP
				Lett.}\ }\textbf {\bibinfo {volume} {33}},\ \bibinfo {pages} {499} (\bibinfo
		{year} {1981})}\BibitemShut {NoStop}%
	\bibitem [{\citenamefont {G.Springholz}(2003)}]{Springholz2003}%
	\BibitemOpen
	\bibfield  {author} {\bibinfo {author} {\bibnamefont {G.Springholz}},\
	}\bibfield  {title} {\enquote {\bibinfo {title} {Molecular beam epitaxy of
				{IV-VI} heterostructures and superlattices},}\ }in\ \href@noop {} {\emph
		{\bibinfo {booktitle} {Lead Chalcogenides: Physics and Applications}}},\
	\bibinfo {editor} {edited by\ \bibinfo {editor} {\bibfnamefont
			{H.}~\bibnamefont {Pascher}}, \bibinfo {editor} {\bibfnamefont
			{G.}~\bibnamefont {Bauer}}, \ and\ \bibinfo {editor} {\bibfnamefont
			{D.}~\bibnamefont {Khokhlov}}}\ (\bibinfo  {publisher} {Taylor \& Francis},\
	\bibinfo {year} {2003})\BibitemShut {NoStop}%
	\bibitem [{\citenamefont {Pfeffer}\ and\ \citenamefont
		{Zawadzki}(1995)}]{Pfeffer1995}%
	\BibitemOpen
	\bibfield  {author} {\bibinfo {author} {\bibfnamefont {P.}~\bibnamefont
			{Pfeffer}}\ and\ \bibinfo {author} {\bibfnamefont {W.}~\bibnamefont
			{Zawadzki}},\ }\bibfield  {title} {\enquote {\bibinfo {title} {{Spin
					splitting of conduction subbands in GaAs-Ga$_{0.7}$Al$_{0.3}$As
					heterostructures}},}\ }\href {\doibase doi:10.1103/PhysRevB.52.R14332}
	{\bibfield  {journal} {\bibinfo  {journal} {Phys. Rev. B}\ }\textbf {\bibinfo
			{volume} {52}},\ \bibinfo {pages} {R14 332--335} (\bibinfo {year}
		{1995})}\BibitemShut {NoStop}%
	\bibitem [{\citenamefont {Dugaev}\ and\ \citenamefont
		{Khmelnitsky}(1984)}]{Dugaev1984}%
	\BibitemOpen
	\bibfield  {author} {\bibinfo {author} {\bibfnamefont {V.K.}\ \bibnamefont
			{Dugaev}}\ and\ \bibinfo {author} {\bibfnamefont {D.E.}\ \bibnamefont
			{Khmelnitsky}},\ }\bibfield  {title} {\enquote {\bibinfo {title}
			{{Magnetoresistance of metal films with low impurity concentrations in a
					parallel magnetic field}},}\ }\href@noop {} {\bibfield  {journal} {\bibinfo
			{journal} {Sov. Phys. JETP}\ }\textbf {\bibinfo {volume} {59}},\ \bibinfo
		{pages} {1038--1041} (\bibinfo {year} {1984})}\BibitemShut {NoStop}%
	\bibitem [{\citenamefont {Beenakker}\ and\ \citenamefont {van
			Houten}(1988)}]{Beenakker1988}%
	\BibitemOpen
	\bibfield  {author} {\bibinfo {author} {\bibfnamefont {C.~W.~J.}\
			\bibnamefont {Beenakker}}\ and\ \bibinfo {author} {\bibfnamefont
			{H.}~\bibnamefont {van Houten}},\ }\bibfield  {title} {\enquote {\bibinfo
			{title} {Boundary scattering and weak localization of electrons in a magnetic
				field},}\ }\href {\doibase 10.1103/PhysRevB.38.3232} {\bibfield  {journal}
		{\bibinfo  {journal} {Phys. Rev. B}\ }\textbf {\bibinfo {volume} {38}},\
		\bibinfo {pages} {3232--3240} (\bibinfo {year} {1988})}\BibitemShut {NoStop}%
	\bibitem [{\citenamefont {Raichev}\ and\ \citenamefont
		{P.Vasilopoulos}(2000)}]{Raichev2000}%
	\BibitemOpen
	\bibfield  {author} {\bibinfo {author} {\bibfnamefont {O.E.}\ \bibnamefont
			{Raichev}}\ and\ \bibinfo {author} {\bibnamefont {P.Vasilopoulos}},\
	}\bibfield  {title} {\enquote {\bibinfo {title} {Weak-localization
				corrections to the conductivity of double quantum wells},}\ }\href
	{http://stacks.iop.org/0953-8984/12/i=5/a=307} {\bibfield  {journal}
		{\bibinfo  {journal} {Journal of Physics: Condensed Matter}\ }\textbf
		{\bibinfo {volume} {12}},\ \bibinfo {pages} {589} (\bibinfo {year}
		{2000})}\BibitemShut {NoStop}%
	\bibitem [{\citenamefont {Sacksteder}\ \emph {et~al.}(2014)\citenamefont
		{Sacksteder}, \citenamefont {Arnardottir}, \citenamefont {Kettemann},\ and\
		\citenamefont {Shelykh}}]{Sacksteder2014}%
	\BibitemOpen
	\bibfield  {author} {\bibinfo {author} {\bibfnamefont {Vincent~E.}\
			\bibnamefont {Sacksteder}}, \bibinfo {author} {\bibfnamefont {Kristin~Bjorg}\
			\bibnamefont {Arnardottir}}, \bibinfo {author} {\bibfnamefont {Stefan}\
			\bibnamefont {Kettemann}}, \ and\ \bibinfo {author} {\bibfnamefont {Ivan~a.}\
			\bibnamefont {Shelykh}},\ }\bibfield  {title} {\enquote {\bibinfo {title}
			{{Topological effects on the magnetoconductivity in topological
					insulators}},}\ }\href {\doibase 10.1103/PhysRevB.90.235148} {\bibfield
		{journal} {\bibinfo  {journal} {Phys. Rev. B}\ }\textbf {\bibinfo {volume}
			{90}},\ \bibinfo {pages} {235148} (\bibinfo {year} {2014})}\BibitemShut
	{NoStop}%
	\bibitem [{\citenamefont {Fukuyama}(1980)}]{Fukuyama1980}%
	\BibitemOpen
	\bibfield  {author} {\bibinfo {author} {\bibfnamefont {Hidetoshi}\
			\bibnamefont {Fukuyama}},\ }\bibfield  {title} {\enquote {\bibinfo {title}
			{Non-metallic behaviors of two-dimensional metals and effect of intervalley
				impurity scattering},}\ }\href {\doibase 10.1143/PTPS.69.220} {\bibfield
		{journal} {\bibinfo  {journal} {Prog. Theor. Phys. Suppl.}\ }\textbf
		{\bibinfo {volume} {69}},\ \bibinfo {pages} {220--231} (\bibinfo {year}
		{1980})}\BibitemShut {NoStop}%
	\bibitem [{\citenamefont {Belitz}\ and\ \citenamefont
		{Das~Sarma}(1987)}]{Belitz1987}%
	\BibitemOpen
	\bibfield  {author} {\bibinfo {author} {\bibfnamefont {D.}~\bibnamefont
			{Belitz}}\ and\ \bibinfo {author} {\bibfnamefont {S.}~\bibnamefont
			{Das~Sarma}},\ }\bibfield  {title} {\enquote {\bibinfo {title} {Inelastic
				phase-coherence time in thin metal films},}\ }\href {\doibase
		10.1103/PhysRevB.36.7701} {\bibfield  {journal} {\bibinfo  {journal} {Phys.
				Rev. B}\ }\textbf {\bibinfo {volume} {36}},\ \bibinfo {pages} {7701--7704}
		(\bibinfo {year} {1987})}\BibitemShut {NoStop}%
	\bibitem [{\citenamefont {Prinz}\ \emph {et~al.}(1999)\citenamefont {Prinz},
		\citenamefont {Brunthaler}, \citenamefont {Ueta}, \citenamefont {Springholz},
		\citenamefont {Bauer}, \citenamefont {Grabecki},\ and\ \citenamefont
		{Dietl}}]{Prinz1999}%
	\BibitemOpen
	\bibfield  {author} {\bibinfo {author} {\bibfnamefont {A.}~\bibnamefont
			{Prinz}}, \bibinfo {author} {\bibfnamefont {G.}~\bibnamefont {Brunthaler}},
		\bibinfo {author} {\bibfnamefont {Y.}~\bibnamefont {Ueta}}, \bibinfo {author}
		{\bibfnamefont {G.}~\bibnamefont {Springholz}}, \bibinfo {author}
		{\bibfnamefont {G.}~\bibnamefont {Bauer}}, \bibinfo {author} {\bibfnamefont
			{G.}~\bibnamefont {Grabecki}}, \ and\ \bibinfo {author} {\bibfnamefont
			{T.}~\bibnamefont {Dietl}},\ }\bibfield  {title} {\enquote {\bibinfo {title}
			{Electron localization in
				{$n\ensuremath{-}{\mathrm{Pb}}_{1\ensuremath{-}x}{\mathrm{Eu}}_{x}\mathrm{Te}$}},}\
	}\href {\doibase 10.1103/PhysRevB.59.12983} {\bibfield  {journal} {\bibinfo
			{journal} {Phys. Rev. B}\ }\textbf {\bibinfo {volume} {59}},\ \bibinfo
		{pages} {12983--12990} (\bibinfo {year} {1999})}\BibitemShut {NoStop}%
	\bibitem [{\citenamefont {Polley}\ \emph {et~al.}(2018)\citenamefont {Polley},
		\citenamefont {Buczko}, \citenamefont {Forsman}, \citenamefont {Dziawa},
		\citenamefont {Szczerbakow}, \citenamefont {Rechci\'nski}, \citenamefont
		{Kowalski}, \citenamefont {Story}, \citenamefont {Trzyna}, \citenamefont
		{Bianchi}, \citenamefont {\v{C}abo}, \citenamefont {Hofmann}, \citenamefont
		{Tjernberg},\ and\ \citenamefont {Balasubramanian}}]{Polley2018}%
	\BibitemOpen
	\bibfield  {author} {\bibinfo {author} {\bibfnamefont {C.~M.}\ \bibnamefont
			{Polley}}, \bibinfo {author} {\bibfnamefont {R.}~\bibnamefont {Buczko}},
		\bibinfo {author} {\bibfnamefont {A.}~\bibnamefont {Forsman}}, \bibinfo
		{author} {\bibfnamefont {P.}~\bibnamefont {Dziawa}}, \bibinfo {author}
		{\bibfnamefont {A.}~\bibnamefont {Szczerbakow}}, \bibinfo {author}
		{\bibfnamefont {R.}~\bibnamefont {Rechci\'nski}}, \bibinfo {author}
		{\bibfnamefont {B.~J.}\ \bibnamefont {Kowalski}}, \bibinfo {author}
		{\bibfnamefont {T.}~\bibnamefont {Story}}, \bibinfo {author} {\bibfnamefont
			{M.a}\ \bibnamefont {Trzyna}}, \bibinfo {author} {\bibfnamefont
			{M.}~\bibnamefont {Bianchi}}, \bibinfo {author} {\bibfnamefont {A.~G.}\
			\bibnamefont {\v{C}abo}}, \bibinfo {author} {\bibfnamefont {P.}~\bibnamefont
			{Hofmann}}, \bibinfo {author} {\bibfnamefont {O.}~\bibnamefont {Tjernberg}},
		\ and\ \bibinfo {author} {\bibfnamefont {T.}~\bibnamefont
			{Balasubramanian}},\ }\bibfield  {title} {\enquote {\bibinfo {title}
			{Fragility of the {Dirac} cone splitting in topological crystalline insulator
				heterostructures},}\ }\href {\doibase 10.1021/acsnano.7b07502} {\bibfield
		{journal} {\bibinfo  {journal} {ACS Nano}\ }\textbf {\bibinfo {volume}
			{12}},\ \bibinfo {pages} {617--626} (\bibinfo {year} {2018})}\BibitemShut
	{NoStop}%
	\bibitem [{\citenamefont {Rechci{\'{n}}ski}\ and\ \citenamefont
		{Buczko}(2018)}]{Rechcinski2018}%
	\BibitemOpen
	\bibfield  {author} {\bibinfo {author} {\bibfnamefont {Rafa{\l}}\
			\bibnamefont {Rechci{\'{n}}ski}}\ and\ \bibinfo {author} {\bibfnamefont
			{Ryszard}\ \bibnamefont {Buczko}},\ }\bibfield  {title} {\enquote {\bibinfo
			{title} {{Topological states on uneven (Pb,Sn)Se (001) surfaces}},}\ }\href
	{\doibase 10.1103/PhysRevB.98.245302} {\bibfield  {journal} {\bibinfo
			{journal} {Phys. Rev. B}\ }\textbf {\bibinfo {volume} {98}},\ \bibinfo
		{pages} {245302} (\bibinfo {year} {2018})}\BibitemShut {NoStop}%
	\bibitem [{\citenamefont {Linder}\ \emph {et~al.}(2009)\citenamefont {Linder},
		\citenamefont {Yokoyama},\ and\ \citenamefont {Sudb\o{}}}]{Linder2009}%
	\BibitemOpen
	\bibfield  {author} {\bibinfo {author} {\bibfnamefont {Jacob}\ \bibnamefont
			{Linder}}, \bibinfo {author} {\bibfnamefont {Takehito}\ \bibnamefont
			{Yokoyama}}, \ and\ \bibinfo {author} {\bibfnamefont {Asle}\ \bibnamefont
			{Sudb\o{}}},\ }\bibfield  {title} {\enquote {\bibinfo {title} {Anomalous
				finite size effects on surface states in the topological insulator
				{${\text{Bi}}_{2}{\text{Se}}_{3}$}},}\ }\href {\doibase
		10.1103/PhysRevB.80.205401} {\bibfield  {journal} {\bibinfo  {journal} {Phys.
				Rev. B}\ }\textbf {\bibinfo {volume} {80}},\ \bibinfo {pages} {205401}
		(\bibinfo {year} {2009})}\BibitemShut {NoStop}%
	\bibitem [{\citenamefont {Lu}\ \emph {et~al.}(2010)\citenamefont {Lu},
		\citenamefont {Shan}, \citenamefont {Yao}, \citenamefont {Niu},\ and\
		\citenamefont {Shen}}]{HaiZhouLu2010}%
	\BibitemOpen
	\bibfield  {author} {\bibinfo {author} {\bibfnamefont {Hai-Zhou}\
			\bibnamefont {Lu}}, \bibinfo {author} {\bibfnamefont {Wen-Yu}\ \bibnamefont
			{Shan}}, \bibinfo {author} {\bibfnamefont {Wang}\ \bibnamefont {Yao}},
		\bibinfo {author} {\bibfnamefont {Qian}\ \bibnamefont {Niu}}, \ and\ \bibinfo
		{author} {\bibfnamefont {Shun-Qing}\ \bibnamefont {Shen}},\ }\bibfield
	{title} {\enquote {\bibinfo {title} {Massive dirac fermions and spin physics
				in an ultrathin film of topological insulator},}\ }\href {\doibase
		10.1103/PhysRevB.81.115407} {\bibfield  {journal} {\bibinfo  {journal} {Phys.
				Rev. B}\ }\textbf {\bibinfo {volume} {81}},\ \bibinfo {pages} {115407}
		(\bibinfo {year} {2010})}\BibitemShut {NoStop}%
	\bibitem [{\citenamefont {Liu}\ \emph {et~al.}(2010)\citenamefont {Liu},
		\citenamefont {Zhang}, \citenamefont {Yan}, \citenamefont {Qi}, \citenamefont
		{Frauenheim}, \citenamefont {Dai}, \citenamefont {Fang},\ and\ \citenamefont
		{Zhang}}]{Zhang2010}%
	\BibitemOpen
	\bibfield  {author} {\bibinfo {author} {\bibfnamefont {Chao-Xing}\
			\bibnamefont {Liu}}, \bibinfo {author} {\bibfnamefont {HaiJun}\ \bibnamefont
			{Zhang}}, \bibinfo {author} {\bibfnamefont {Binghai}\ \bibnamefont {Yan}},
		\bibinfo {author} {\bibfnamefont {Xiao-Liang}\ \bibnamefont {Qi}}, \bibinfo
		{author} {\bibfnamefont {Thomas}\ \bibnamefont {Frauenheim}}, \bibinfo
		{author} {\bibfnamefont {Xi}~\bibnamefont {Dai}}, \bibinfo {author}
		{\bibfnamefont {Zhong}\ \bibnamefont {Fang}}, \ and\ \bibinfo {author}
		{\bibfnamefont {Shou-Cheng}\ \bibnamefont {Zhang}},\ }\bibfield  {title}
	{\enquote {\bibinfo {title} {Oscillatory crossover from two-dimensional to
				three-dimensional topological insulators},}\ }\href {\doibase
		10.1103/PhysRevB.81.041307} {\bibfield  {journal} {\bibinfo  {journal} {Phys.
				Rev. B}\ }\textbf {\bibinfo {volume} {81}},\ \bibinfo {pages} {041307}
		(\bibinfo {year} {2010})}\BibitemShut {NoStop}%
\end{thebibliography}
%

\end{document}